\input{aipcheck}

\documentclass[
    ,final            % use final for the camera ready runs
  ]
  {aipproc}

\layoutstyle{6x9}

\begin{document}

\title{Does an Average White Dwarf Have Enough Mass to Prevent Accretion Disk Tilt?}


\author{M.M. Montgomery}{
  address={Department of Physics, University of Central Florida, Orlando, FL  32816, USA}
}

\begin{abstract}
In a recent publication, we introduce the lift force as a common source to accretion disk tilt that is likely relevant to accretion disk systems.  Lift is generated by slightly different supersonic gas stream speeds flowing over and under the disk at the bright spot.   In this conference proceeding, we focus on whether the average white dwarf has enough mass to prevent a disk tilt in non-magnetic Cataclysmic Variables (CVs) with accretion disks.  Assuming a white dwarf mass of 0.6M$_{\odot}$ and a disk mass of 10$^{-11}$M$_{\odot}$, we vary the secondary mass to establish theoretical minimum mass transfer rates needed to induce and maintain a disk tilt of four degrees around the line of nodes.  For mass ratios in the range \( (0.13 \le q=M_{2}M^{-1} \le 0.45) \), we confirm that the secondary mass does not contribute significantly to disk tilt.   We also confirm that the average white dwarf does not have enough mass to prevent a disk tilt.  We find that disk tilt may be likely in low mass transfer rate systems such as CV SU UMa's.
\end{abstract}

\maketitle

\section{Introduction}
Over the last forty years or so, cyclic brightness modulations that have periods longer than orbital periods have been observed in a variety of systems including Cataclysmic Variables (CVs).  Warner (2003) suggests that long period modulations can be taken as indirect evidence of a tilted disk.  Montgomery (2009a) shows that tilted disks could retrogradely precess due to the net tidal torque by the secondary on a misaligned accretion disk, like Earth's retrograde precession due to the net tidal torque by the Moon and Sun on the equatorial bulge of the spinning and tilted Earth.  In Montgomery \& Martin (2010) we introduce a source to accretion disk tilt that depends on the mass of the central compact object, the surface area of the accretion disk, and a slight variation of the mass transfer rate over and under the accretion disk at the bright spot.  Using the analytical expression developed in that work, we establish gross magnitudes of the minimum, average mass transfer rates needed to induce and maintain a disk tilt of four degrees around a typical white dwarf mass primary.   In the following sections, we generate analytical data that we later analyze.  We also provide some discussions and conclusions.   

\section{Gross Magnitude Estimates of $\dot{M}$ (kg s$^{-1}$)}
In Montgomery \& Martin (2010), we establish an analytical expression for the gross magnitude of the minimum mass transfer rate to induce a disk tilt, 
\begin{eqnarray}
|\dot{M}|  & \ge & \frac{32G \Sigma m M_{1}}{9r_{d}^{2} |v_{o}| (1-\beta^{2})}  \left( \frac{b}{r_{d}}\right)^{2}\sin\theta + \\ \nonumber
   &    &  \frac{3r_{d}GmM_{2}}{2|v_{o}|(1-\beta^{2})(d^{2}+\frac{9}{16}r_{d}^{2}-\frac{3}{2}r_{d}d\cos\theta)^{3/2}}  \left( \frac{b}{r_{d}}\right)^{2}\sin\theta.
\end{eqnarray}

\noindent
In this equation, $\dot{M}$ is the mass transfer rate, $G$ is the universal gravitational constant, $\Sigma m$ is the sum of gas particles in the disk (i.e., the total mass of the disk, not including the mass of the primary),  $M_{1}$ is the mass of the white dwarf primary, $b$ is the radius of the gas stream just prior to striking the bright spot, $\theta$ is the obliquity angle, $r_{d}$ is the radius of the disk, $\bf{v_{o}}$ is the gas stream velocity over the accretion disk, $\beta$ is a fraction, $M_{2}$ is the secondary mass, and $d$ is distance between the primary and secondary.  Note that this expression does not include effects such as gas compressibility, viscosity, and gas density.  Experiments are needed to establish these parameters as well as coefficients of drag and lift as discussed in Montgomery \& Martin (2010) and thus inclusion of these unknowns is outside the scope of this work.    

In this work, we assume an average primary mass $M_{1}$=0.6M$_{\odot}$ as recent studies show that the majority of white dwarf masses lie between $ (0.5 < M_{1}$ M$_{\odot}^{-1} < 0.7)  $  (Kepler et al., 2007).  We also assume a gas particle mass $m=2\times10^{14}$kg and $\Sigma m$=10$^{-11}$M$_{\odot}$ (Montgomery 2009b), minimum obliquity angle $\theta=4^{o}$ for negative superhumps to potentially be observed in light curves (Montgomery 2009b), approximate supersonic gas stream speed over the accretion disk $\bf{v_{o}}$=5$\times10^{5}$ms$^{-1}$ (Montgomery \& Martin, 2010), and 10\% variation in gas stream speeds flowing over the disk relative to under at the bright spot or $\beta=0.9$ (Montgomery \& Martin, 2010).  The radius of the disk is dependent on the mass ratio $q$ and can be found from Paczynski (1977), \( r_{d}=\frac{0.6d}{1+q}, \) which is valid for mass ratios in the range \( (0.03 < q < 1) \).  To find the separation distance, we assume the secondary mass-radius relation, \( R_{2}=(M_{2}/M_{\odot})^{13/15} R_{\odot}, \) which applies for (0.08 $\le M_{2}$M$_{\odot}^{-1} \le 1.0) $ as discussed in Warner (2003). In this relation, $R_{2}$ is in solar radii $R_{\odot}$.  To find the separation distance, we can substitute the radius of the secondary into the Eggleton (1983) relation \( \frac{R_{2}}{d} = \frac{0.49 q^{2/3}}{0.6 q^{2/3} + ln(1 + q^{1/3})}, \) a relation that is good for all mass ratios, accurate to better than 1\%.  Then, the orbital period can be found from Newton's version of Kepler's Third Law, \( P^{2} = \frac{4\pi^{2}}{G(M_{1}+M_{2})}d^{3} \) where $P$ is the orbital period.  The remaining unknown is the width of the gas stream.  Warner (2003) lists an effective cross-section of the stream as \( Q \approx 2.4\times10^{17} \left( \frac{T_{s}}{10^{4}} \right) P^{2} \) where $Q=\pi b^{2}$ is in cm$^{2}$, surface temperature of the secondary $T_{s}$ is in Kelvin, and orbital period is in hours.  This equation is based on the isothermal sound velocity which assumes  an atomic hydrogen environment.  If we assume a dM4 companion, the typical average secondary in a DA binary (Silvestri et al. 2006), then from Bessel (1995), $T_{eff} \approx T_{s}$=3130K. Table 1 lists the theoretical data obtained from the above equations for mass ratios in the range \( (0.13 \le q \le 0.45) \).  Expected mass transfer rates in various CV systems are listed in Table 2.  Note in Table 1 that the mass transfer rate due to the $M_{1}$ term refers to the first term on the right hand side of Equation (1).  Likewise, the mass transfer rate due to the $M_{2}$ term refers to the second term on the right hand side of Equation (1).

 %%%%%%%%%%%%%%%%%TABLE 1%%%%%%%%%%
\begin{table}
  \begin{tabular}{@{}lcccccccccc@{}}
  \hline
  q	& d  &$r_{d}$ & $P_{orb}$ &b$\times10^{6}$ & \multicolumn{2}{c}{Min $|\dot{M}|$ (kgs$^{-1}$)} & &  & \multicolumn{2}{c}{Min $|\dot{M}|$ ($M_{\odot}$yr$^{-1}$)} \\
		\cline{6-7} \cline{10-11}
 &  ($R_{\odot}$) &  ($R_{\odot}$) &  (hr) & (m) &  $M_{1}$ Term & $M_{2}$ Term && & $M_{1}$ Term & $M_{2}$ Term \\
 \hline
0.13   &  0.49    &  0.26   & 1.16  &   1.79 & 1.2 $\times10^{13}$ & 4.6$\times10^{6}$ & & & 1.9$\times10^{-10}$ & 7.3$\times10^{-17}$ \\
0.15   &  0.53    &  0.28   & 1.30  &   2.01 & 1.2 $\times10^{13}$ & 4.7$\times10^{6}$ & & & 1.9$\times10^{-10}$ & 7.5$\times10^{-17}$ \\
0.20   &  0.63    &  0.32   & 1.64  &   2.54 & 1.2 $\times10^{13}$ & 4.9$\times10^{6}$ & & & 1.9$\times10^{-10}$ & 7.8$\times10^{-17}$ \\
0.25   &  0.72    &  0.35   & 1.96  &   3.04 & 1.2 $\times10^{13}$ & 5.1$\times10^{6}$ & & & 1.9$\times10^{-10}$ & 8.0$\times10^{-17}$ \\
0.30   &   0.81   &  0.37   & 2.27  &   3.51 & 1.2 $\times10^{13}$ & 5.1$\times10^{6}$ & & & 1.9$\times10^{-10}$ & 8.0$\times10^{-17}$ \\
0.35   &  0.88    &  0.39   & 2.56  &   3.95 & 1.2 $\times10^{13}$ & 5.1$\times10^{6}$ & & & 1.9$\times10^{-10}$ & 8.0$\times10^{-17}$ \\
0.40   &   0.96   &  0.41   & 2.84  &   4.38 & 1.2 $\times10^{13}$ & 5.1$\times10^{6}$ & & & 1.9$\times10^{-10}$ & 8.0$\times10^{-17}$ \\
0.45   &  1.03    &  0.43   & 3.10  &   4.80 & 1.3 $\times10^{13}$ & 5.1$\times10^{6}$ & & & 2.0$\times10^{-10}$ & 8.0$\times10^{-17}$ \\
\hline
\end{tabular}
\caption{Theoretical Data for M$_{1}$=0.6M$_{\odot}$}
\label{tab:a}
\end{table}
%%%%%%%%%%%%%%%%%%%%%%%%%%%%%%%%%%%

 %%%%%%%%%%%%%%%%%TABLE 2%%%%%%%%%%
\begin{table}
  \begin{tabular}{@{}lclc@{}}
  \hline 
CV Name &  P$_{orb}$ (hr) & Min $|\dot{M}|$ (kgs$^{-1}$) & Ref \\
 \hline
ER UMa   &  1.52    &  4 $\times10^{13}$       & Warner (2003), Hellier (2001)\\
SU UMa   &  1.83    & 9.9 $\times10^{12}$     & Hellier (2001) \\
WZ  Sge   &  1.36    &  2 $\times10^{12}$       & Warner (2003) \\
Z Cam      &   6.96   &  6 $\times10^{13}$       & Warner (2003) \\
U Gem     &   4.25   &  3.2 $\times10^{13}$    & Warner (2003) \\
VY Scl      &   3.99   &  6 $\times10^{14}$       & Warner (2003) \\
SW Sex   &   3.24   &  6 $\times10^{14}$        & Warner (2003) \\
OY Car    &   1.52   &  4 $\times10^{12}$        & Warner (2003) \\
Z Cha      &    1.79  &   5 $\times10^{12}$       & Smak (2004) \\
IP Peg     &    3.80  &  1.4 $\times10^{12}$     & Warner (2003) \\
YZ Cnc    &    2.10  &  2.38 $\times10^{12}$  & Smak (2004) \\
VW Hyi    &    1.78  & 1 $\times10^{12}$         & Smak (2004) \\
\hline
\end{tabular}
\caption{Mass Transfer Rates in Various CV Systems}
\label{tab:a}
\end{table}
%%%%%%%%%%%%%%%%%%%%%%%%%%%%%%%%%%%

\section{Analysis of the Data and Discussion}
As shown in Table 1, the red dwarf has no effect on the minimum mass transfer rate needed to cause a tilt in the disk.  In other words, the second term on the right hand side of Equation (1) can be ignored which agrees with a conclusion drawn in Montgomery \& Martin (2010).  

Also shown in Table 1, the highest average gross magnitude of the minimum mass transfer rate needed to induce and maintain a disk tilt of four degrees is $
\sim10^{13}$kgs$^{-1}$ or $\sim10^{-10}$M$_{\odot}$yr$^{-1}$.  This value is lower than known mass transfer rates in CVs that have observed negative superhumps in their light curves and/or observed retrograde precessional periods (see e.g., Montgomery 2009b).  This value is also lower than or near the listed mass transfer rates in ER UMa, SU UMa, Z Cam, U Gem, VY Scl, and SW Sex in Table 2.  Note, mass transfer rates for SU UMa, WZ Sge, OY Car, Z Cha, YZ Cnc, and VW Hyi in Table 2 are estimated when these systems are in quiescence.  If these systems are entering into an outburst, when mass transfer rates are expected to be higher, then disk tilt is more likely.  These results suggest that low mass transfer rate systems such as SU UMa's may exhibit negative superhumps and/or long period modulations, results consistent with those suggested for MN Dra by Pavlenko et al. (2010).  

As we do not know the available surface area of truncated disks in IP systems, we did not calculate gross magnitudes of mass transfer rates needed to induce a disk tilt in these systems.  However, disk tilt may be possible in these systems as well as in U Gem's.

\section{Conclusions}

In this work, we establish the minimum mass transfer rates needed to induce and maintain a disk tilt of four degrees for various non-magnetic CV systems that have an average white dwarf primary mass of 0.6$M_{\odot}$.  We confirm the results of Montgomery \& Martin (2010).  In addition, we predict that many CVs with accretion disks and average 0.6$M_{\odot}$ mass primaries could have tilted disks if the supersonic gas speed flowing over the disk varies slightly from that flowing under the disk.  

\begin{theacknowledgments}
We would like to acknowledge the international travel grant of the American Astronomical Society and the 17th European White Dwarf Workshop for partial financial support to present this work.  
\end{theacknowledgments}

\bibliographystyle{aipproc}   % if natbib is available

%\bibliography{sample}

\IfFileExists{\jobname.bbl}{}
 {\typeout{}
  \typeout{******************************************}
  \typeout{** Please run "bibtex \jobname" to optain}
  \typeout{** the bibliography and then re-run LaTeX}
  \typeout{** twice to fix the references!}
  \typeout{******************************************}
  \typeout{}
 }

\end{document}